# Fast algorithm to calculate density of states

R. E. Belardinelli and V. D. Pereyra*

*Departamento de Física, Laboratorio de Ciencias de Superficie, Universidad Nacional de San Luis, CONICET, Chacabuco 917, 5700 San Luis, Argentina*

An algorithm to calculate the density of states, based on the well-known Wang-Landau method, is introduced. Independent random walks are performed in different restricted ranges of energy, and the resultant density of states is modified by a function of time, $F(t) \propto t^{-1}$, for large time. As a consequence, the calculated density of state, $g_m(E, t)$, approaches asymptotically the exact value $g_{ex}(E)$ as $\propto t^{-1/2}$, avoiding the saturation of the error. It is also shown that the growth of the interface of the energy histogram belongs to the random deposition universality class.

## I. INTRODUCTION

The Wang-Landau (WL) algorithm [1] has been one of the most interesting and refreshing improvements in the Monte Carlo (MC) simulation scheme in the last decade, applying to a broad spectrum of interesting problems in statistical physics and biophysics [1–9].

The method is based in an algorithm to calculate the density of states $g(E)$—i.e., the number of all possible states (or configurations) for an energy level $E$ of the system. In that way, thermodynamic observables, including free energy over a wide range of temperature, can be calculated with one single simulation.

Instead, most conventional Monte Carlo algorithms such as Metropolis importance sampling [10], Swendsen-Wang cluster flipping [11], etc. [12], generate a canonical distribution $g(E)e^{-E/k_BT}$ at a given temperature. Such distributions are so narrow that, with conventional Monte Carlo simulations, multiple runs are required to determine thermodynamic quantities over significant ranges of temperatures.

It is important to note that the multicanonical ensemble method [13–16] proposed by Berg *et al.* estimates also the density of states $g(E)$ first, then performs a random walk with a flat histogram in the desired region in the phase space. This method has been proven to be very efficient in studying first-order phase transitions where simple canonical simulations have difficulty in overcoming the tunneling barrier between coexisting phases at the transition temperature [13,16–23]. However, in multicanonical simulations, the density of states need not necessarily be very accurate, as long as the simulation generates a relatively flat histogram and overcomes the tunneling barrier in energy space. This is because the subsequent reweighting [13,15] does not depend on the accuracy of the density of states as long as the histogram can cover all important energy levels with sufficient statistics.

Since Wang and Landau introduced the multiple-range random walk algorithm to calculate the density of states (DOS), there have been numerous proposed improvements

───────
* rbelar@unsl.edu.ar and vpereyra@unsl.edu.ar

[24–31] and studies of the efficiency and convergence of this algorithm [26,30]. However, there are limitations of the method which remain still unsolved, such as, for example, the behavior of the tunneling time, which is a bound for the performance of any flat-histogram algorithm, as is discussed in Ref. [30], where it is shown that it limits the convergence in the WL algorithm. Other important unanswered questions related particularly with the WL method are as follows: (i) How is the flatness of the histogram related to the accuracy? (ii) What is the relation between the modification factor and the error? (iii) Is there some relation between the refinement parameter and the stopping condition that increases the efficiency? (iv) Is there any universality behavior related to this algorithm? In this paper an algorithm based in the Wang-Landau method is introduced. The main goal of the proposed algorithm is that the refinement parameter needs to be scaled down as $1/t$.

The remainder of this paper is arranged as follows. In Sec. II, the Wang-Landau algorithm and its dynamical behavior are discussed. In Sec. III, the algorithm is introduced in detail and some applications are discussed. Finally, the conclusions are given in Sec. IV.

## II. WANG-LANDAU ALGORITHM

In the Wang-Landau algorithm (WL), an initial energy range of interest is identified, $E_{min} \leqslant E \leqslant E_{max}$, and a random walk is performed in this range. During the random walk, two histograms are updated: one for the density of states, $g(E)$, which represents the current or running estimate, and one for visits to distinct energy states, $H(E)$. Before the simulation begins $H(E)$ is set to zero and $g(E)$ is set to unity. The random walk is performed by choosing an initial state $i$ in the energy range $E_{min} \leqslant E \leqslant E_{max}$. Trial moves are then attempted and moves are accepted according to the transition probability

$$p(E_i \rightarrow E_f) = \min\left(1, \frac{g(E_i)}{g(E_f)}\right), \qquad (1)$$

where $E_i$ ($E_f$) is the initial (final) state, respectively, and $p(E_i \rightarrow E_f)$ is the transition probability from the energy level $E_i$ to $E_f$. Whenever a trial move is accepted, a histogram entry corresponding to $n$ is incremented according to $H(E_n) = H(E_n) + 1$ and $g(E_n) = g(E_n)f$, where $f$ is an arbitrary convergence factor, which is generally initialized as $e$. If a move is rejected, the new configuration is discarded and the histogram entry corresponding to the old configuration is incremented according to $H(E_0) = H(E_0) + 1$; at the same time, the density of states is incremented according to $g(E_0) = g(E_0)f$. This process is repeated until the energy histogram becomes sufficiently flat. When that happens, the energy histogram is reset to zero and the convergence factor is decreased according to $f_{k+1} = \sqrt{f_k}$, where $f_k$ is the convergence factor corresponding to stage $k$. The process is continued until $f$ becomes sufficiently close to 1 [say, $f < \exp(10^{-8})$].

In practice, the relation $S(E) = \log[g(E)]$ is generally used in order to fit all possible values of $g(E)$ into double-precision numbers.

In order to understand the WL method and its limitations, let us describe the time behavior of $\langle H(t)\rangle = \frac{1}{N}\Sigma_E H(E,t)$, where $N$ is the number of states of different energies and $H(E,t)$ stands for the mean height of the histogram in $E$ at time $t$ (with $t = j/N$, where $j$ is the number of trial moves attempted) which is the MC time for the rest of the paper. Note that $t$ is normalized to the entire range of energy, $N$, and not to the number of lattice sites, $L^2$.

The wide histogram is defined as $\Delta H(t) = H_{max}(t) - H_{min}(t)$, where $H_{max}(t)$ and $H_{min}(t)$ are the maximum and minimum values of $H$ at time $t$, respectively.

Other quantities of interest are $\frac{\Delta H}{\langle H\rangle}$, the function $F = \log[f]$, and the errors $\epsilon(E,t)$ and $\eta(E,t)$, defined as

$$\epsilon(E,t) = \left| 1 - \frac{\log[g_n(E,t)]}{\log[g_{ex}(E)]} \right| \tag{2}$$

and

$$\eta(E,t) = \left| 1 - \frac{g_n(E,t)}{g_{ex}(E)} \right|, \tag{3}$$

where $g_n(E,t)$ and $g_{ex}(E,t)$ are the experimental and the exact values of the DOS [$g_n(E,t)$ is normalized with respect to the exact DOS at the ground state]. Then, the mean values are $\langle\epsilon(t)\rangle = \frac{1}{(N-1)}\Sigma_E \epsilon(E,t)$ and $\langle\eta(t)\rangle = \frac{1}{(N-1)}\Sigma_E \eta(E,t)$.

The dynamics of the WL algorithm can be analyzed observing the behavior of these quantities, as shown in Figs. 1(a) and 1(b), where $\langle H\rangle$, $\Delta H$, $\frac{\Delta H}{\langle H\rangle}$, $F$, $\langle\epsilon(t)\rangle$, and $\langle\eta(t)\rangle$ are plotted as a function of MC time for the Ising model in a two-dimensional square lattice with $L = 8$, over the whole range of energies, $N = 63$. Note that in this case, $t = \frac{L^2}{N}t'$, where $t'$ is normalized to the number of the lattice sites, $L^2$. The exact value of the density of states, $g_{ex}(E)$, is obtained from the method developed in Ref. [32], and $g_n(E,t)$ is averaged over 256 independent samples.

For $0 < t < t_c$, $F_0 = \text{const}$ [i.e. $F_0 = \log(e) = 0.434\ 294\ 4 \ldots$], where $t_c$ is the number of MC trial flips between adjacent records in the histogram, typically $t_c = 10000\frac{L^2}{N}$ MC step; both errors $\langle\epsilon(t)\rangle$ and $\langle\eta(t)\rangle$ are very large; the average height of the histogram grows linearly with time as $\langle H(t)\rangle = t$; $\Delta H$ follows the same behavior of $\langle H\rangle$, while $H_{min} = 0$; as soon as $H_{min} \gtrsim 0$, $\Delta H$ remains constant and consequently the rate $\frac{\Delta H}{\langle H\rangle}$ begins to decrease inversely with time, $\frac{\Delta H}{\langle H\rangle} \propto t^{-1}$. When this parameter fulfills the flatness criteria (i.e., 95%, horizontal line in the figure) and the Monte Carlo time is the adequate $t > t_c$, then $F$ is modified to a new value ($F_{k+1} = F_k/2$), the histogram $H(E)$ is reset, and the errors $\langle\epsilon(t)\rangle$ and $\langle\eta(t)\rangle$ are reduced, remaining constant until the parameter $F$ is modified again. New reductions of $F$ lead to lower values of $\langle\epsilon(t)\rangle$ and $\langle\eta(t)\rangle$ and the error curves go down.

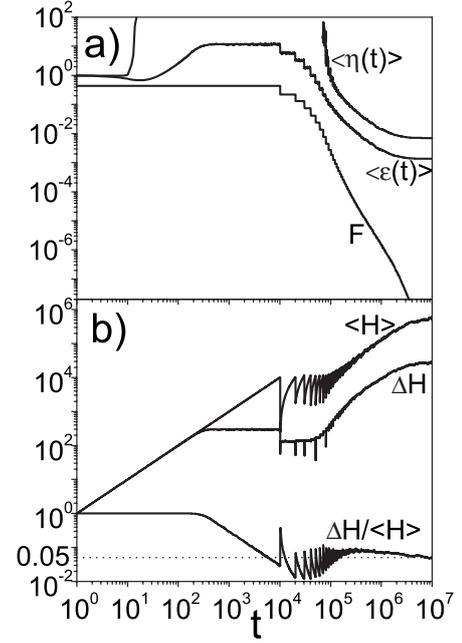

FIG. 1. Dynamical behavior of (a) $F$, $\langle\epsilon(t)\rangle$, $\langle\eta(t)\rangle$ and (b) $\langle H\rangle$, $\Delta H$, $\frac{\Delta H}{\langle H\rangle}$ for the Ising model in a two-dimensional square lattice with $L = 8$ using the Wang-Landau method. The quantities are obtained averaging over 256 independent samples. The flatness criterion was 95%.

However, after a certain time both errors reach a saturation value. It turns out that $\langle H\rangle$, $\Delta H$, and $\frac{\Delta H}{\langle H\rangle}$ are also saturated. It seems that the saturation in the error is due to the fact that the parameter $F$ is reduced in an inadequate way, every time that the histogram becomes flat.

In the WL algorithm, any function $F$ may be used for the modification factor, as long as it decreases monotonically to zero. One of the ways of reducing $F$ is to use the following assignment relation: $F_{(k+1)} = \frac{F_k}{m}$ with $m = 2$. One could suggest to change the value of $m$ to optimize the error; however, other values of $m$ present the same behavior, saturating in all cases as is shown in Fig. 2. In order to render the WL method efficient, $m$ should be variable in each interval controlled by the flatness in the histogram.

## III. $t^{-1}$ TIME-DEPENDENT ALGORITHM

In this paper an algorithm to obtain the density of states is introduced by determining an appropriate law for the temporary change of $F$ in such a way that the saturation in the errors is avoided. One proceeds in the same way as in the WL method, identifying the range of the energy of interest, performing the random walk in this range, and using the acceptance probability defined in Eq. (1). The algorithm is as follows:

(i) Choose an initial configuration and set $S(E) = 0$, $H(E) = 0$ for all $E$ and $F_0 = 1$ and also fix $F_{final}$.

(ii) The system changes from $E_i$ to $E_f$ according to the probability given in Eq. (1).

(iii) Increment the histogram and the density of state, $H(E) \rightarrow H(E) + 1$, and $S(E) \rightarrow S(E) + F_k$.

(iv) After some fixed sweeps (i.e., 1000 MC time) check $H(E)$; if $H(E) \neq 0$ for all $E$, then refine $F_{k+1} = F_k/2$ and reset the histogram $H(E)$.

(v) If $F_{k+1} \leq t^{-1}$, then $F_{k+1} = F(t) = t^{-1}$ and in what follows $F(t)$ not ι

(vi) If $F(t) < F_{final}$, then the process is stopped. Otherwise go to (ii).

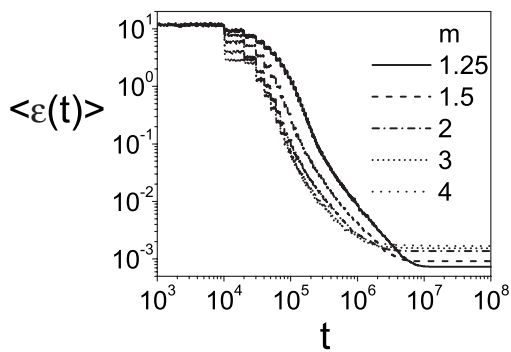

FIG. 2. Dynamical behavior of $\langle \epsilon(t) \rangle$ for different values of the parameter $m$. The flatness criterion was 95%. As one can see, in all cases the error saturates.

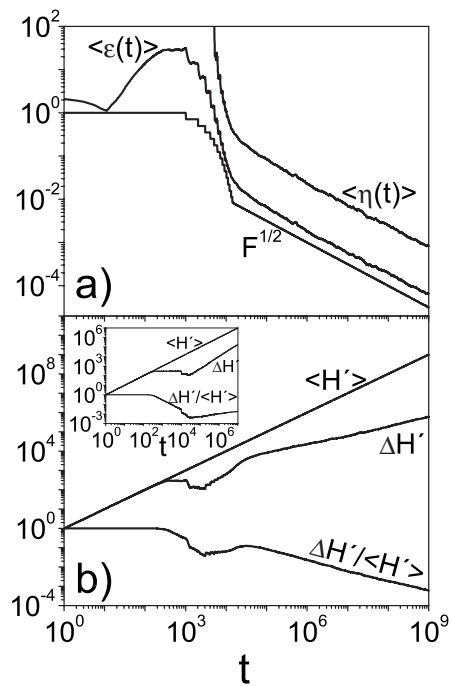

FIG. 3. Dynamical behavior of (a) $\sqrt{F}$, $\langle \epsilon(t) \rangle$, $\langle \eta(t) \rangle$ and (b) $\langle H' \rangle$, $\Delta H'$, $\frac{\Delta H'}{\langle H' \rangle}$, for the Ising model in a 2D square lattice with $L = 8$ using the algorithm introduced here. The quantities are obtained averaging over 88 independent samples. In a double-logarithmic scale, the data points for the quantities $\Delta H'$, $\frac{\Delta H'}{\langle H' \rangle}$, $\langle \eta(t) \rangle$, and $\langle \epsilon(t) \rangle$ fall on a straight line with the values of the slopes equal to $0.518 \pm 0.015$, $-0.482 \pm 0.015$, $-0.486 \pm 0.008$, and $-0.488 \pm 0.008$,

The main goal of the algorithm proposed here is to find an appropriate variation of the refinement parameter as a function of time, as is given in (v).

To analyze the dynamic of the proposed algorithm it is convenient to introduce a new histogram $H'(E)$ analogous to $H(E)$, which is not reset in the whole experiment.

In Fig. 3 one can observe a test of the algorithm through the time behavior of the quantities (a) $\langle \epsilon(t) \rangle$, $\langle \eta(t) \rangle$, and $\sqrt{F}$, and (b) $\langle H'(t) \rangle$, $\Delta H'$, and $\frac{\Delta H'}{\langle H' \rangle}$, for the same system as that described in Fig. 1. One can see that the errors $\langle \eta(t) \rangle$ and $\langle \epsilon(t) \rangle$ are proportional to $\frac{1}{\sqrt{t}}$ and $F(t) = t^{-1}$ for large $t$. On the other hand, $\langle H'(t) \rangle = t$ for all time $t$ (this is due to the fact that its quantity is not reset during the whole experiment), $\Delta H' \propto \sqrt{t}$ and $\frac{\Delta H'}{\langle H' \rangle} \propto \frac{1}{\sqrt{t}}$.

In the inset of the Fig. 3(b), the behavior of the nonreset histogram $\langle H'(t) \rangle$, $\Delta H'$, and $\frac{\Delta H'}{\langle H' \rangle}$ for the WL algorithm is included. Clearly, it is observed that both quantities $\langle H'(t) \rangle$ and $\Delta H'$ are proportional to $t$ and $\frac{\Delta H'}{\langle H' \rangle} \rightarrow$ const for large $t$.

One can verify that the errors $\langle \eta(t) \rangle$ and $\langle \epsilon(t) \rangle$ go as $\propto \sqrt{F(t)}$; this is the main characteristic of the algorithm. On the other hand, the fluctuation of the histogram goes as $\Delta H \propto \frac{1}{\sqrt{F(t)}}$; this relation was also obtained by different authors in Refs. [26,31] for the WL method. Based on the time behavior of the histogram growth and its fluctuation, one can conclude that it belongs to the universality class of the random deposition model [33].

The election of $F(t) = 1/\langle H'(t) \rangle$ is not arbitrary; in fact, if $F(t)$ decreases more slowly than $1/\langle H'(t) \rangle$, the algorithm will lose unnecessary time to arrive at the same precision in $g(E)$ [this can be easily tested, changing the exponent in the function $F(t)$ and observing the behavior of the errors]. On the contrary, if $F(t)$ decreases faster than $1/\langle H'(t) \rangle$, the error saturates, as happens in the WL method. $F(t) = t^{-1}$ is a limit function.

To compare the speed and convergence of the algorithm proposed here with the WL method, the error $\langle \epsilon(t) \rangle$ versus time $t$ is plotted in Fig. 4 for both cases. It is observed that for any value of the flatness condition in the WL method, the error saturates. Taking the value of the error at the saturation time for each curve, one can observe the $t^{-1/2}$ dependence; such behavior suggests the functionality of $F(t)$. One can also observe that the error calculated by the method proposed here, at a given time $t$, is always smaller than the one corresponding to WL with any value of the flatness condition. On the other hand, the time that the WL algorithm takes is al-ways greater than that corresponding to our algorithm.

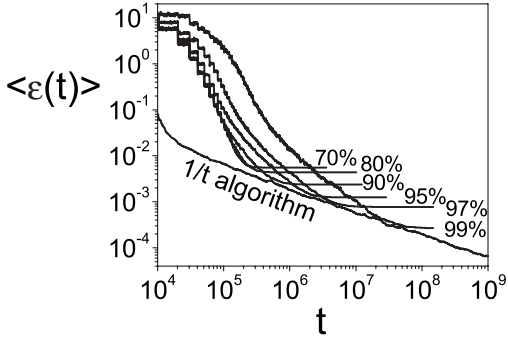

FIG. 4. Comparison between the error $\langle\epsilon(t)\rangle$ calculated using the WL method with different flatness condition and that obtained by our algorithm for the Ising model in a 2D square lattice with $L=8$.

Figure 5 shows the errors versus time for one interval of energy, $E/L^2 \in [-2,0]$, for an Ising model with $L=50$ and 80 averages. In the same figure, $\sqrt{F(t)}$ and $\frac{\Delta H'}{\langle H'\rangle}$ are also included. In the inset, one can see $\epsilon(E)$ and $\eta(E)$ as a function of $\frac{E}{L^2}$ at the final time $t_{final}$ for a typical independent sample.

As is expected, in the Ising model one can compare the exact value of the energy density function, at least for size $L \leqslant 50$, with the experimental one. However, the error and $\frac{\Delta H'}{\langle H'\rangle}$ have the same time functionality. This fact is necessary when the exact energy density function is not previously known, as in other statistical models. Then, observing the behavior of $\frac{\Delta H'}{\langle H'\rangle}$ as a function of time, one can obtain information about the accuracy and convergence of the algorithm.

To check this fact, in Fig. 6 the critical temperature $T_c$ is shown as a function of the size of the lattice for a ten-state Potts model. $T_c$ has been obtained using the method of the double peak described in Ref. [34]. The size of the lattices goes from $L=40$ to $L=120$. To compare both methods, the results given by the WL algorithm (Ref. [1]) are also shown. As one can see, the precision of the proposed algorithm is rather better than that obtained by the WL algorithm [in the same conditions, $f_{final}=\exp(10^{-8})$]. The critical temperature, obtained by using our algorithm according to finite-size theory, is $T_c=0.701\,237\pm0.000\,008$ (the exact result for the infinite system is $T_c=0.701\,232\ldots$; see Ref. [35]).

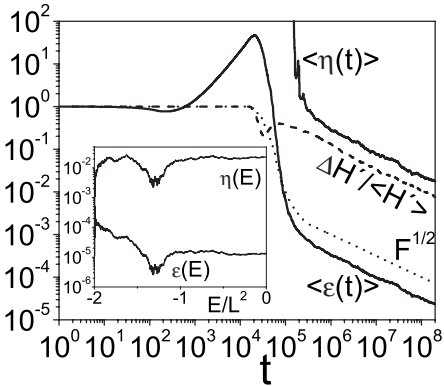

FIG. 5. Time behavior of $\sqrt{F}$, $\frac{\Delta H'}{\langle H'\rangle}$, $\langle\eta(t)\rangle$, and $\langle\epsilon(t)\rangle$ for the whole energy range $E/L^2 \in [-2,0]$ for an Ising model in a 2D square lattice with $L=50$ and $N=1250$ averaging over 80 independent samples. In a double-logarithmic scale, the data points for the quantities $\frac{\Delta H'}{\langle H'\rangle}$, $\langle\eta(t)\rangle$, and $\langle\epsilon(t)\rangle$ fall on a straight line with the values of the slopes equal to $-0.51\pm0.03$, $-0.51\pm0.03$, and $-0.54\pm0.05$, respectively. In the inset the behavior is shown of the errors $\eta$ and $\epsilon$ as a function of the energy $E/L^2$ at the final stage of the simulation ($2\times10^8$ MC steps) for a typical independent sample.

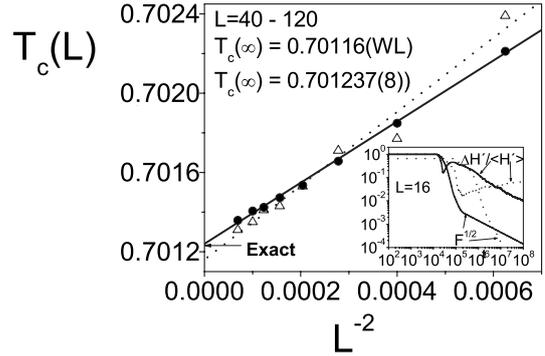

FIG. 6. Extrapolation of finite lattice "transition temperatures" for the algorithm proposed here (solid circles) and WL algorithm (open triangles) for a 2D $Q=10$ Potts model. In the inset the time behavior of $\frac{\Delta H'}{\langle H'\rangle}$ and $\sqrt{F}$ is shown for both the WL (dashed lines) and our algorithm (solid lines), with $L=16$.

In the inset of the figure, the time behavior of $\frac{\Delta H'}{\langle H'\rangle}$ and $\sqrt{F(t)}$ is shown for the WL and the algorithm proposed here, for $L=16$. While in the first case, $\frac{\Delta H'}{\langle H'\rangle}$ goes to a constant value, even for very low values of $\sqrt{F(t)}$, in the second one, $\frac{\Delta H'}{\langle H'\rangle} \propto \sqrt{F(t)} \propto t^{-1/2}$.

It is important to note that in our method, as in the WL method, during the random walk especially in the early time, the algorithm does not satisfy the detailed balance condition exactly, because $g(E)$ is modified constantly during the random walk. After the refinement parameter assumes the $t^{-1}$ functionality, the random walk algorithm satisfies the detailed balance with accuracy proportional to $\frac{1}{t}$.

## IV. CONCLUSIONS

Summarizing, in this paper an accurate algorithm to calculate the energy density function is proposed. The algorithm considers that the refinement parameter $F$ in the well-known WL method is a function of time, which decreases at the same speed as the average height in the energy histogram increases. In this way, the density of states calculated, $g_m(E,t)$, approaches asymptotically the exact value $g_{ex}(E)$ as $\propto t^{-1/2}$. This simple law avoids the saturation in the error.

Although the function $\frac{\Delta H'}{\langle H'\rangle}$ follows the same time behavior as the error, the histogram is not indispensable in the implementation of the method. In other words, there is a direct relation between the accuracy of the method and the flatness in the histogram, as is shown in this work. Note that in the present algorithm the energy histogram growth belongs to the universality class of the random deposition model, independently of the statistical model that is simulated.

The most important property of the algorithm is the present relation between the error and the function $F(t)$; this characteristic allows the assignment of the final stopping condition to $F(t)$, making it as small as one wants. In this way, the total time of simulation is previously determined for a given $F_{final}$, which is not possible to do with the WL algorithm and its variations.

In conclusion, the proposed algorithm is more efficient, accurate, and easy to implement than other well-known methods.

**ACKNOWLEDGMENTS**

We thank Professor G. Zgrablich for reading the manuscript. This work is partially supported by the CONICET (Argentina).